\documentclass[journal]{IEEEtran}
\usepackage{graphicx}
\usepackage{cite}
\usepackage{amsmath,amssymb,amsfonts}
\usepackage{algorithmic}
\usepackage{textcomp}
\usepackage{booktabs}
\usepackage{xspace}
\usepackage{svg}
\usepackage{amsmath}
\usepackage{footnote}
\usepackage{makecell}
\usepackage{multirow}
\usepackage{enumitem}
\usepackage{array}
\usepackage[nolist,nohyperlinks]{acronym}
\usepackage{hyperref}
\usepackage{array}
\newcolumntype{P}[1]{>{\centering\arraybackslash}p{#1}}
\usepackage{subcaption}
\makesavenoteenv{tabular}
\usepackage{caption, subcaption}
\usepackage{csquotes}
\usepackage[acronym]{glossaries}

\acrodef{mIoU}{mean Intersection over Union}
\acrodef{CADx}{Computer-Aided diagnosis}
\acrodef{CNN}{Convolutional Neural Network}
\acrodef{FPS}{Frame Per Second}
\acrodef{WCE}{Wireless capsule endoscopy}
\acrodef{AI}{Artificial Intelligence}
\acrodef{VCE}{Video capsule endoscopy}
\acrodef{DSC}{Dice Coefficient}
\acrodef{SOTA}{state-of-the-art}
\acrodef{MCC}{Matthews correlation coefficient}
\acrodef{GI}{gastrointestinal tract}
\begin{document}


\title{Video Capsule Endoscopy Classification using Focal Modulation Guided Convolutional Neural Network}

\author{
\IEEEauthorblockN{Abhishek Srivastava\IEEEauthorrefmark{1}, Nikhil Kumar Tomar\IEEEauthorrefmark{2},
Ulas Bagci\IEEEauthorrefmark{3},
Debesh Jha\IEEEauthorrefmark{3}
}\\
\IEEEauthorblockA{\IEEEauthorrefmark{1} Computer Vision and Pattern Recognition Unit, Indian Statistical Institute \\
\IEEEauthorrefmark{2}School of Informatics and Computer Science, Indira Gandhi National Open University\\ 
\IEEEauthorrefmark{3} Department of Radiology, Feinberg School of Medicine, Northwestern University, USA\\ 
}
}

\maketitle
\begin{abstract}
Video capsule endoscopy is a hot topic in computer vision and medicine. Deep learning can have a positive impact on the future of video capsule endoscopy technology. It can improve the anomaly detection rate, reduce physicians' time for screening, and aid in real-world clinical analysis. \ac{CADx} classification system for video capsule endoscopy has shown a great promise for further improvement. For example, detection of cancerous polyp and bleeding can lead to swift medical response and improve the survival rate of the patients. To this end, an automated CADx system must have high throughput and decent accuracy. In this paper, we propose \textit{FocalConvNet}, a focal modulation network integrated with lightweight convolutional layers for the classification of small bowel anatomical landmarks and luminal findings. FocalConvNet leverages focal modulation to attain global context and allows global-local spatial interactions throughout the forward pass. Moreover, the convolutional block with its intrinsic inductive/learning bias and capacity to extract hierarchical features allows our FocalConvNet to achieve favourable results with high throughput. We compare our FocalConvNet with other \ac{SOTA} on Kvasir-Capsule, a large-scale VCE dataset with 44,228 frames with 13 classes of different anomalies. Our proposed method achieves the weighted F1-score, recall and \ac{MCC} of 0.6734, 0.6373 and 0.2974, respectively outperforming other \ac{SOTA} methodologies. Furthermore, we report the highest throughput of 148.02 images/second rate to establish the potential of FocalConvNet in a real-time clinical environment. The code of the proposed FocalConvNet is available at \url{https://github.com/NoviceMAn-prog/FocalConvNet}.
\end{abstract}

\begin{IEEEkeywords}
Video capsule endoscopy, small intestine,  deep learning, Kvasir-Capsule, gastrointestinal image classification
\end{IEEEkeywords}
\IEEEpeerreviewmaketitle

\section{Introduction}
\begin{figure}[!t]
    \centering
    \includegraphics [width=4cm]{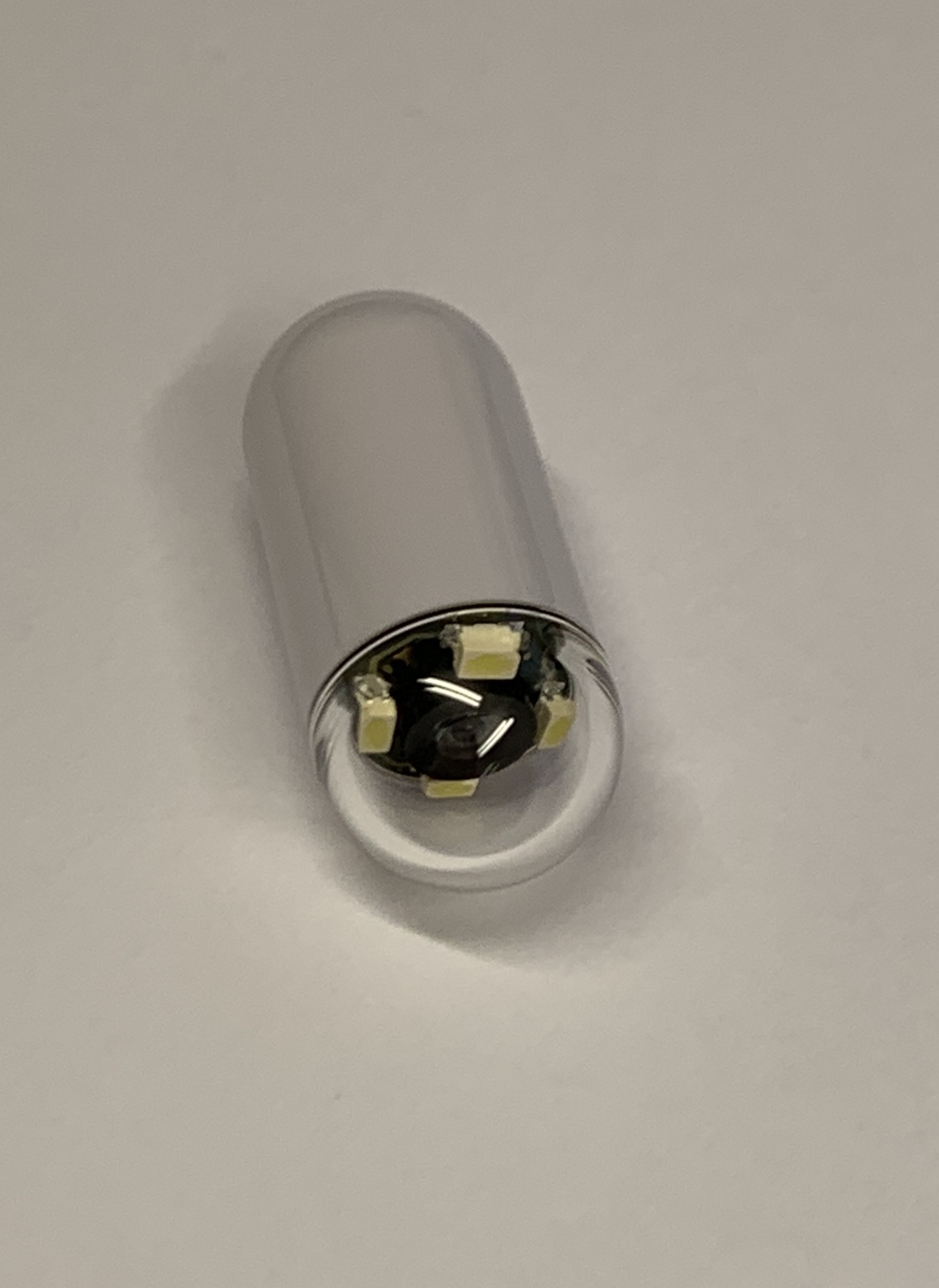}
    \caption{Olympus EC-S10 endocapsule~\cite{smedsrud2021kvasir}}.
    \label{fig:capsule}
    \vspace{-10mm}
\end{figure}

\acf{WCE} is a technology that allows gastroenterologists to visualize the small bowel (intestine).  It was devised by a group of researchers in Baltimore in 1989 and afterwards introduced by Given Imaging Ltd., Yoqneam, Israel, as a commercial instrument~\cite{iddan2000wireless}. The device received FDA approval for use in 2001. \ac{VCE} is useful for differentiating between small intestine bleeding, inflammatory bowel disease and other small bowel abnormalities~\cite{lee2019natural}.  The capsule-shaped pill (Figure~\ref{fig:capsule}) can be swallowed by the patients in the presence of clinical experts without any discomfort~\cite{suman2017detection}. Unlike conventional endoscopy procedures, this procedure investigates the small bowel without pain, sedation and air insufflation.  An additional advantage of \ac{VCE} is that it is non-invasive and easy procedure but plays a crucial role in examination and diagnosis small bowel lesions~\cite{lee2019natural}.%


The small intestine located inside the \ac{GI} tract performs the powerful function of absorbing nutrients~\cite{meerveld2017gastrointestinal}. Ailments caused inside the small intestine can cause grave problems like growth retardation or nutrient deficiencies~\cite{meerveld2017gastrointestinal}. A major obstacle in the detection and subsequent treatment of these diseases is the anatomical location of the small intestine. While imaging of the upper \ac{GI} and large intestine is feasible with endoscopy based methods, the small intestine requires VCE. Figure~\ref{fig:capsule} shows the Olympus EC-S10 endocapsule which was used in Olympus Endocapsule 10 system~\cite{ENDOCAPSULE-10} for capturing videos in the Kvasir-Capsule~\cite{smedsrud2021kvasir} dataset. \ac{VCE} is performed by a small capsule equipped with a wide-angle camera. The patient consumes the capsule and the capsule traverses throughout the \ac{GI} tract and records the video which is later evaluated by a clinical expert. Given the large number of frames generated by the procedure, the subsequent analysis is monotonous and susceptible to human error which can potentially lead to a high-miss rate~\cite{kaminski2010quality}.

\ac{CADx} systems can save significant time and resources involved in \ac{VCE} analysis. In this study, we aim to develop an automated \ac{VCE} classification algorithm that can classify \ac{VCE} frames with high accuracy and real-time speed to be used in the clinical practise. To achieve this, we propose \enquote{\textit{FocalConvNet}} which integrates \ac{CNN} based modules and Focal modulation~\cite{yang2022focal} to establish a new baseline on Kvasir-Capsule~\cite{smedsrud2021kvasir}.


The main contributions of this work are summarized as follows:
\begin{enumerate}
 \item We proposed a novel deep learning architecture, named FocalConvNet, to classify anatomical and luminal findings in video frames of capsule endoscopy. We leverage the focal modulation strategy of aggregating multi-scale context for modulating the query. The focal modulation mechanism is integrated with lightweight convolutional to outperform several \ac{SOTA} image classification methods.
 
 \item FocalConvNet not only achieves the best F1-score and \ac{MCC} in classification on Kvasir-Capsule, but also its lightweight and computationally economical architecture report the highest throughput of 148 images per second.
 
 \item To advance the work on long-tailed anatomical, pathological, and mucosal view classification in \ac{VCE} frames, we provide additional comparisons with \ac{SOTA} \ac{CNN} and transformer based classification networks which otherwise lacked comparable studies.
\end{enumerate}

\section{Related work}
\label{sec:relatedwork}

\subsection{Image classification using convolutional neural networks}
The introduction of AlexNet~\cite{krizhevsky2012imagenet} led convolutional neural networks to be the most prevalent architectures for nearly all computer vision tasks. Since then various architectural advancements have been made. VGG~\cite{simonyan2014very} increased the effective depth in \ac{CNN}s. ResNet~\cite{he2016deep} introduced residual connections in \ac{CNN}. DenseNet~\cite{huang2017densely} proposed densely connected convolutional structures. Multi-scale fusion was introduced in HRNet~\cite{wang2020deep} and further studied in~\cite{9662196,srivastava2021gmsrf}. Spatial and channel-wise attention~\cite{hu2018squeeze,woo2018cbam} further enhanced the performance of existing \ac{CNN}s. Additionally, alterations have been made in convolution operation in depth-wise convolution layer~\cite{chollet2017xception} and point-wise convolution~\cite{hua2018pointwise} layer. Recently, methods like ConvNeXt~\cite{liu2022convnet} and ConvMixer~\cite{trockman2022patches} have integrated some aspects of vision transformers to further increase the performance of \ac{CNN}s.
\subsection{Image classification using vision transformers}
ViT~\cite{dosovitskiy2020image} introduced transformers for vision problems. Since then, this design of architecture has drawn great attention and many incarnations of ViT have been introduced. Swin Transformer used shifting windows for constraining self-attention computation to non-overlapping windows. DeiT~\cite{touvron2021training} relied on a token-based distillation strategy to train data efficient transformers. Further work has been done to reduce the quadratic complexity of the self-attention heads. PoolFormer~\cite{yu2021metaformer} replaced the attention layer with a pooling operation to achieve favourable results. Focal networks~\cite{yang2022focal} replaced self-attention with modulation of the query by aggregating global and local contexts. Although vision transformers have demonstrated proficiency in learning global representations, \ac{CNN}s have several desirable properties. Thus, various methodologies blending convolution and self-attention mechanisms have been put forth recently. ConViT~\cite{d2021convit} used gated positional encoding which combined self-attention with a soft convolutional inductive bias. Although such hybrid networks have been studied in the past, in this work, we combine the efficacy of focal modulation with computationally economical convolutions to introduce a faster and more accurate network for classifying anatomical landmarks, pathological findings, and quality of the mucosal view. Additional work has been done on further reducing architecture complexity and increasing efficiency by; employing sparse attention~\cite{arnab2021vivit,dong2021cswin,rao2021dynamicvit}, Pyramidal designs~\cite{wang2021pyramid,fan2021multiscale}, integration with \ac{CNN}s~\cite{dai2021coatnet,d2021convit}. More information on the advancements in vision transformers can be found in~\cite{khan2021transformers}.

\section{Network architecture}
\begin{figure*}[!t]
    \centering
    \includegraphics[width=0.95\textwidth]{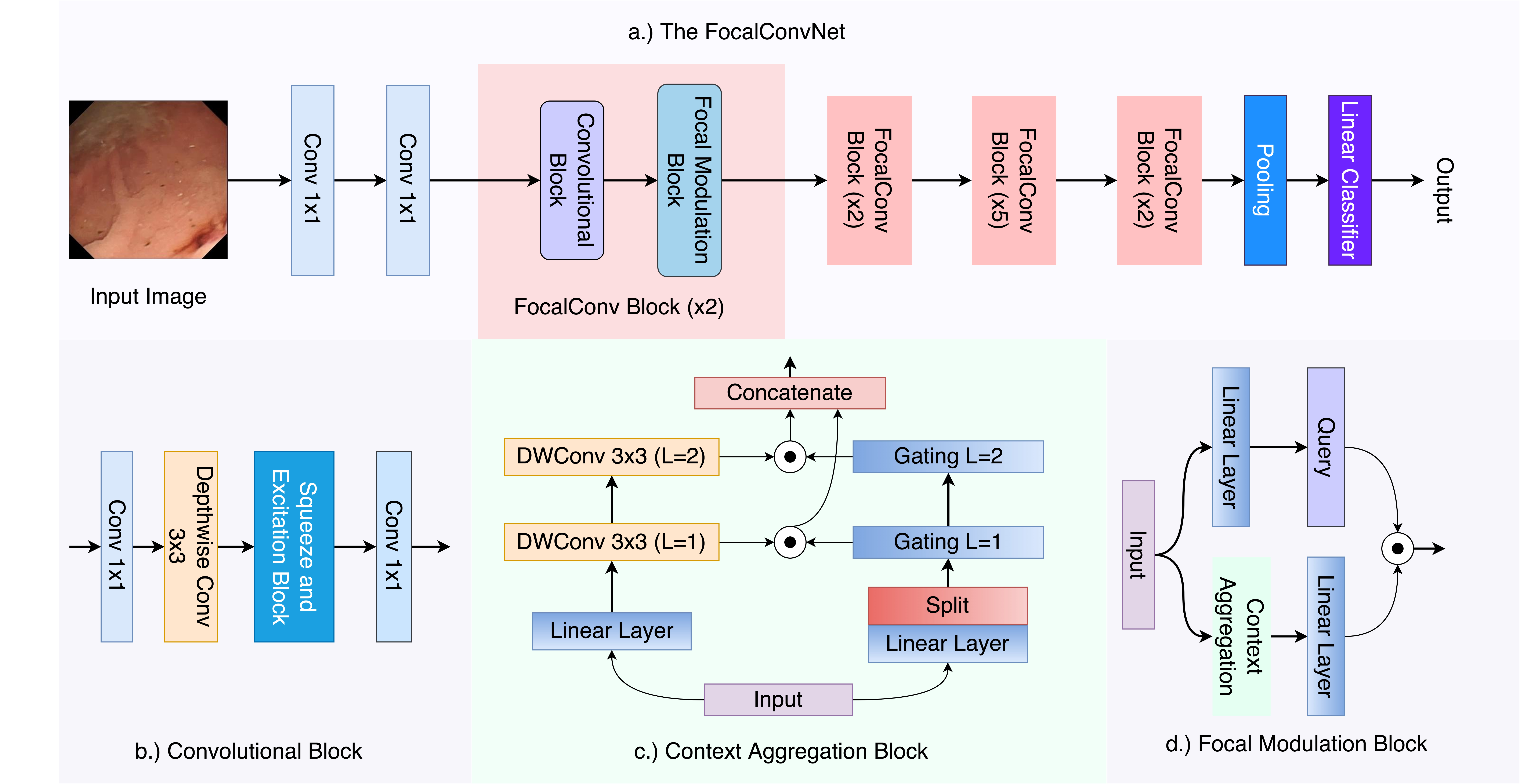}
    \caption{\textbf{The proposed FocalConvNet architectures.} {\bf a}) The complete FocalConvNet architecture (x$K$ represents that the block in question has been used $K$ times in succession),  {\bf b}) The convolutional block used with depth-wise and point-wise convolutions, {\bf c}) Context aggregation sub-module (the figure demonstrates context aggregation when number of focal levels is 2, and can be extended in a similar manner if deeper focal levels are used), {\bf d}) Focal modulation mechanism.}
    \label{fig:FocalConvNet}
    \vspace{-5mm}
\end{figure*}

The architecture of FocalConvNet consists of three steps, initial convolutional stem, FocalConv blocks and the final linear classifier. FocalConv blocks incorporate alternate convolutional and focal modulation blocks (see Figure~\ref{fig:FocalConvNet}(b) and Figure~\ref{fig:FocalConvNet}(d)). Multi-scale context can be gathered and used for modulating the input query through focal modulation blocks. Thus, modelling input dependent global interactions is not as computationally expensive as self-attention strategies utilized by other prominent vision transformers~\cite{yang2022focal}. Keeping with the theme of reducing the computational complexity, our convolutional layer uses depth-wise separable convolutions which significantly reduces the required computational resources as compared to standard convolutions. Blending the two architectures allows us to leverage global multi-scale context while extracting hierarchical features using convolutional layers and retain the advantageous properties of \ac{CNN}s (scale and translation invariance, inductive bias) and transformers (increased generalizability).

\subsection{Convolutional Block}
Figure~\ref{fig:FocalConvNet}(b) illustrates the design of our convolutional block. First, the input feature maps are passed through a convolutional layer with a kernel size of $1$ (point-wise convolutional layer). Hereafter, a depth-wise convolutional layer with kernel size $3$ is used. The ensuing Squeeze and excitation ($\text{S\&E}$) block increases the network's representative power by computing the inter-dependencies between channels. Again, a point-wise convolution operation is used, to keep the model parametrically cheap while retaining the power of a standard convolution operator with kernel size $3$. The output from this block is fed into the Focal Modulation block.
\subsection{Context Aggregation and Focal Modulation}
\label{section:CA}
The input feature map $X_{f}$ is first operated upon by a linear layer to convert it into a new feature space(see Equation~\ref{eq:1}).
\begin{equation}
\label{eq:1}
    M_{0} = Linear(X_{f})
\end{equation}
Let the number of focal levels be $n$. For the $i$'th focal level the output $M_{i}$ is calculated as:
\begin{equation}
\label{eq:2}
    M_{i} = GeLU(DepthConv(M_{i-1}))
\end{equation}
Here, $DepthConv$ is a depth-wise convolutional layer followed by GeLU activation layer. Again depth-wise convolutional layer serves in modeling long range interplay. At each layer $l$ the receptive field can be calculated by $r^{l} = 1 + \sum_{i=1}^{l}(k^l - 1)$. Where, $k^l$ denotes the kernel size for layer $l$. For capturing complete global scale context, average pooling is used to calculate the final ($n+1$)'th feature map, $M_{n+1} = Pool(M_{n})$
Hence, the depth-wise convolutional layer is used to obtain hierarchical features which are used for assembling multi-scale context. Next, a gating mechanism is used to adaptively limit the information propagated by each feature map $M_{i}$ for each focal level $i$. A learnable linear layer is used for obtaining gating weights $G = Linear(X_{f})$, which has the same dimensions as feature vector $M$. Finally $M^{out}$ is obtained by Equation~\ref{eq:3}
\begin{equation}
\label{eq:3}
    M^{out} = \sum_{i=1}^{i=n+1} G_{i} \odot M_{i}
\end{equation}
Here, $\odot$ represents element-wise multiplication. Hereafter, aggregation is performed across channels using $M^{out} = f(M^{out})$, where $f$ is a linear layer. Subsequent focal modulation is performed by the interaction of query($q$) with the modulator $M^{out}$ as $y = q \odot M^{out}$, where $y$ is the output feature vector of the focal modulation block.

\subsection{The FocalConvNet architecture}
Initially, the input image is processed by two convolutional layers with a kernel size of $3$, where the first and the second layer have a stride of $2$ and $1$, respectively. Next the extracted features are fed into 4 consecutive sequences of FocalConv blocks (see Figure~\ref{fig:FocalConvNet}(a)). The aforementioned four sequences consist of 2, 2, 5, and 2 FocalConv blocks, respectively. Each sequence has a focal level of $3$ and reduces the spatial dimension of the input feature vector by a factor of $2$. Finally, an adaptive average pooling layer and a linear classification layer are used to obtain the final prediction.

\begin{figure} [!t]
    \centering
    \includegraphics [width =0.5\textwidth]{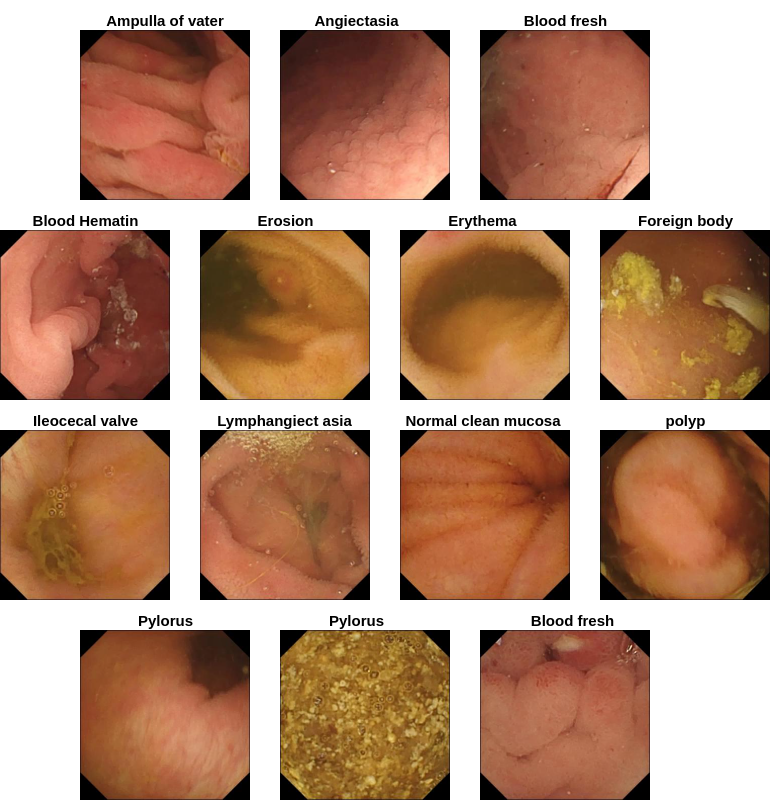}
    \caption{Example images from anatomy and luminal finding classes from Kvasir-Capsule dataset~\cite{jha2022machine}}
    \label{fig:capsuleexample}
    \vspace{-5mm}
\end{figure}

\section{Experimental Setup}

\begin{table*}[!t]
\centering
\footnotesize
\caption{Result comparison on Kvasir-Capsule~\cite{smedsrud2021kvasir}}
\begin{tabular}{@{}l|l|l|l|l|l|l|l|l@{}}
\toprule
\multirow{2}{*}{\textbf{Method}} & \multicolumn{3}{|c|}{\bf Macro Average} & \multicolumn{3}{|c|}{\bf Weighted Average} & \multirow{2}{*}{\textbf{Accuracy}} & \multirow{2}{*}{\textbf{MCC}} \\ 
& \textbf{Precision} & \textbf{Recall} &  \textbf{F1-Score} & \textbf{Precision} & \textbf{Recall} &  \textbf{F1-Score} \\ \hline
GMSRF-Net~\cite{srivastava2021gmsrf} & 0.1568 & 0.1980 & 0.1575 & 0.7431 & 0.6095 & 0.6636 & 0.6090 & 0.2665  \\ \hline
ResNet-152~\cite{he2016deep} & 0.1563 & 0.2049 & 0.1463 & 0.7295 & 0.4886 & 0.5675 & 0.4886 & 0.1979 \\ \hline
DenseNet-169~\cite{huang2017densely} & 0.1884 & 0.2262 & 0.1483 & 0.7141 & 0.5540 & 0.6083 & 0.5540 & 0.2041 \\ \hline
ConvMix-768/32~\cite{trockman2022patches} & 0.1426 & 0.1779 & 0.1188 &  0.7464 & 0.3475 & 0.4428 & 0.3475 & 0.1570 \\ \hline
ConvMix-1536/20~\cite{trockman2022patches} & 0.1722 & 0.2275 & 0.1697 & 0.7431 & 0.6021 & 0.6524 & 0.6021 & 0.2717 \\ \hline
EfficientNetV2-S~\cite{tan2021efficientnetv2} & 0.1623 & 0.2424 & 0.1686 & 0.7563 & 0.5588 & 0.6312 & 0.5588 & 0.2615  \\ \hline
EfficientNetV2-M~\cite{tan2021efficientnetv2} & 0.1558 & 0.2102 & 0.1456 & 0.7347 & 0.5887 & 0.5199 & 0.5199 & 0.2267 \\ \hline 
ConvNeXt-S~\cite{liu2022convnet} & 0.1177 & 0.2310 & 0.1012 & 0.7277 & 0.4349 & 0.5173 & 0.4349 & 0.1758 \\ \hline
ConvNeXt-B~\cite{liu2022convnet} & 0.1311 & 0.1169 & 0.1108 & 0.7276 & 0.3917 & 0.4965 & 0.3918 & 0.1387 \\ \hline
Swin-S~\cite{dong2021cswin} & 0.1538 & 0.2388 & 0.1525 & 0.7390 & 0.5800 & 0.6334 & 0.5800 & 0.2565 \\ \hline
Swin-B~\cite{dong2021cswin} & 0.1496 & 0.2310 & 0.1525 & 0.7134 & 0.5905 & 0.6355 &  0.5905 & 0.2288 \\ \hline
ConViT-S~\cite{d2021convit} & 0.1765 & 0.2182 & 0.1689 & \textbf{0.7673} & 0.5610 & 0.6312 & 0.5610 & 0.2769 \\ \hline
ConViT-B~\cite{d2021convit} & 0.1769 & 0.2534 & 0.1700 & 0.7406 & 0.5541 & 0.6160 & 0.5541 & 0.2443 \\ \hline 
Focal-S~\cite{yang2022focal} & 0.1403 & 0.1919 & 0.1344 & 0.7352 & 0.4883 & 0.5690 & 0.4883 & 0.2060 \\ \hline
Focal-B~\cite{yang2022focal} & 0.1394 & 0.1869 & 0.1368 & 0.7300 & 0.5110 & 0.5873 & 0.5110 & 0.2092 \\ \hline
FocalConvNet(Ours) & \textbf{0.2438} & \textbf{0.2745} & \textbf{0.2178} & 0.7557 & \textbf{0.6373} & \textbf{0.6734} & \textbf{0.6373} & \textbf{0.2964} \\ 

\bottomrule
\end{tabular}
\label{tab:result1}
\end{table*}

In this section, we will provide details about the dataset, evaluation metrics, implementation details, and data augmentation techniques used. 

\subsection{Dataset}
We select Kvasir-Capsule~\cite{smedsrud2021kvasir}, the world's largest video capsule endoscopy dataset for experimentation. The Kvasir-Capsule comprises of 44,228 labelled images from 13 classes of anatomical and luminal findings. Figure~\ref{fig:capsuleexample} shows the examples images from the Kvasir-Capsule dataset. The number of images per class is as followed; Normal mucosa - 34,606; Reduced Mucosal View - 2399; Pylorus - 1520; Polyp - 64; Lymphoid Hyperplasia - 592; Ileo-Cecal valve - 1417; Hematin - 12; Foreign Bodies - 776; Erythematous - 238; Erosion - 438; Blood - 446; Angiectasia - 866; Ulcer - 854. It can be observed that the number of images in class \enquote{Normal Mucosa} is significantly large as compared to any other class and hence, the dataset is heavily imbalanced. As done by the authors in~\cite{smedsrud2021kvasir}, we remove the \enquote{Polyp} and \enquote{Hematin} classes as the number of images is very scarce in these classes. 

\subsection{Implementation details}
We use 23,061 images for training and 24,092 for testing. Random horizontal, vertical flipping and random rotation were used to augment data. Images are resized to $224\times224$ before being fed into the models. The implementation of our proposed FocalConvNet is done using the PyTorch framework. SGD optimizer with a momentum of $0.9$ and a learning rate of $0.001$ is used during training. We use author-released source code for training and each method is trained from scratch. Throughput is calculated on a single Tesla V100 GPU with a batch size of $6$. We have used weighted categorical cross-entropy as our loss function. Each experiment is performed on NVIDIA DGX-2 machine that uses NVIDIA V100 Tensor core GPUs.

\subsection{Evaluation metrics}
For evaluation purposes, we have chosen standard computer vision metrics such as F1-score, precision, recall, accuracy, throughput, and \ac{MCC}. \ac{MCC} and F1-score are the most preferred metric for the class imbalance problem. Detailed information about the evaluation metrics can be found in~\cite{jha2021comprehensive,jha2022machine}.  

\begin{figure*}[!t]
    \centering
    \includegraphics[width=1\textwidth]{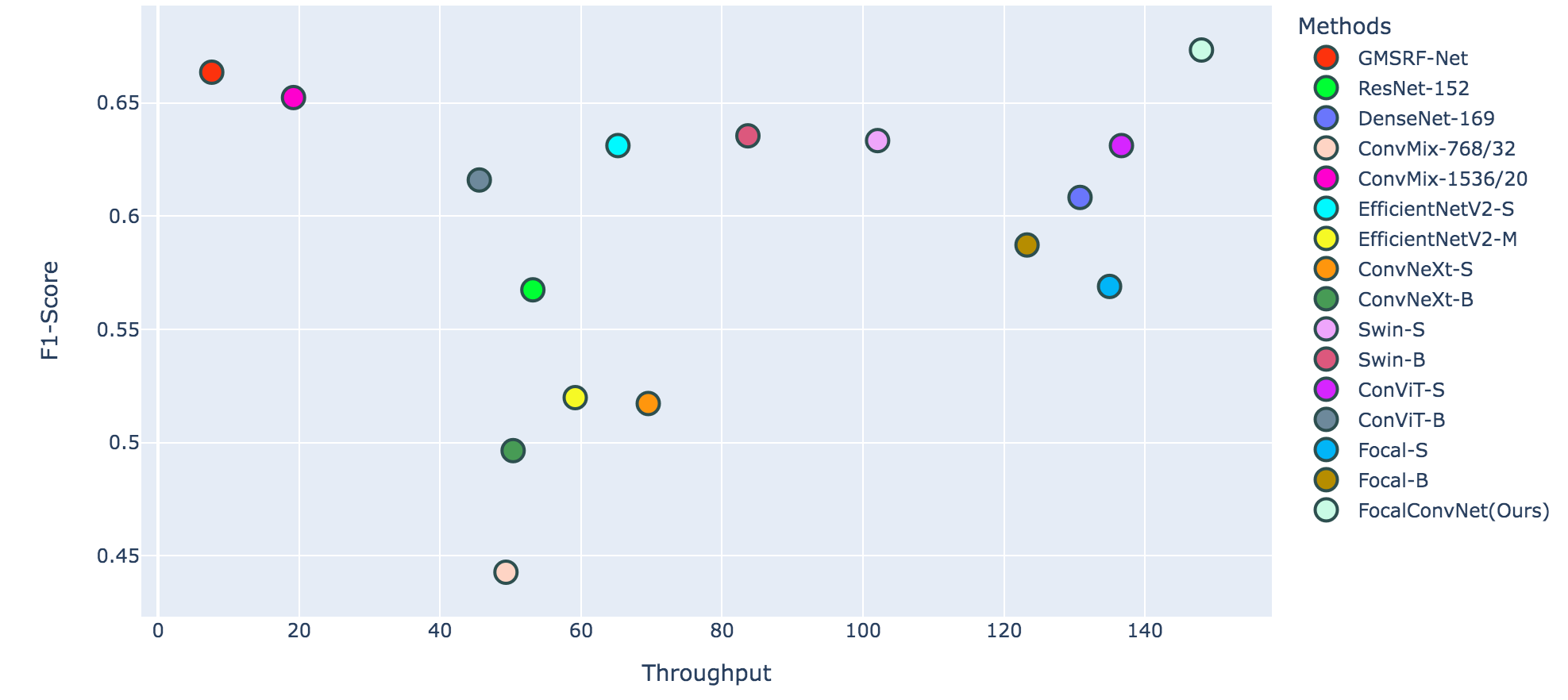}
    \caption{\textbf{Plot of F1-score vs throughput for each method}. We demonstrate that FocalConvNet is capable of achieving the best weighted F1-score while maintaining a high throughput.}
    \label{fig:f1vsthrough}
    \vspace{-5mm}
\end{figure*}

\begin{figure*}[!t]
    \centering
    \includegraphics[width=1\textwidth]{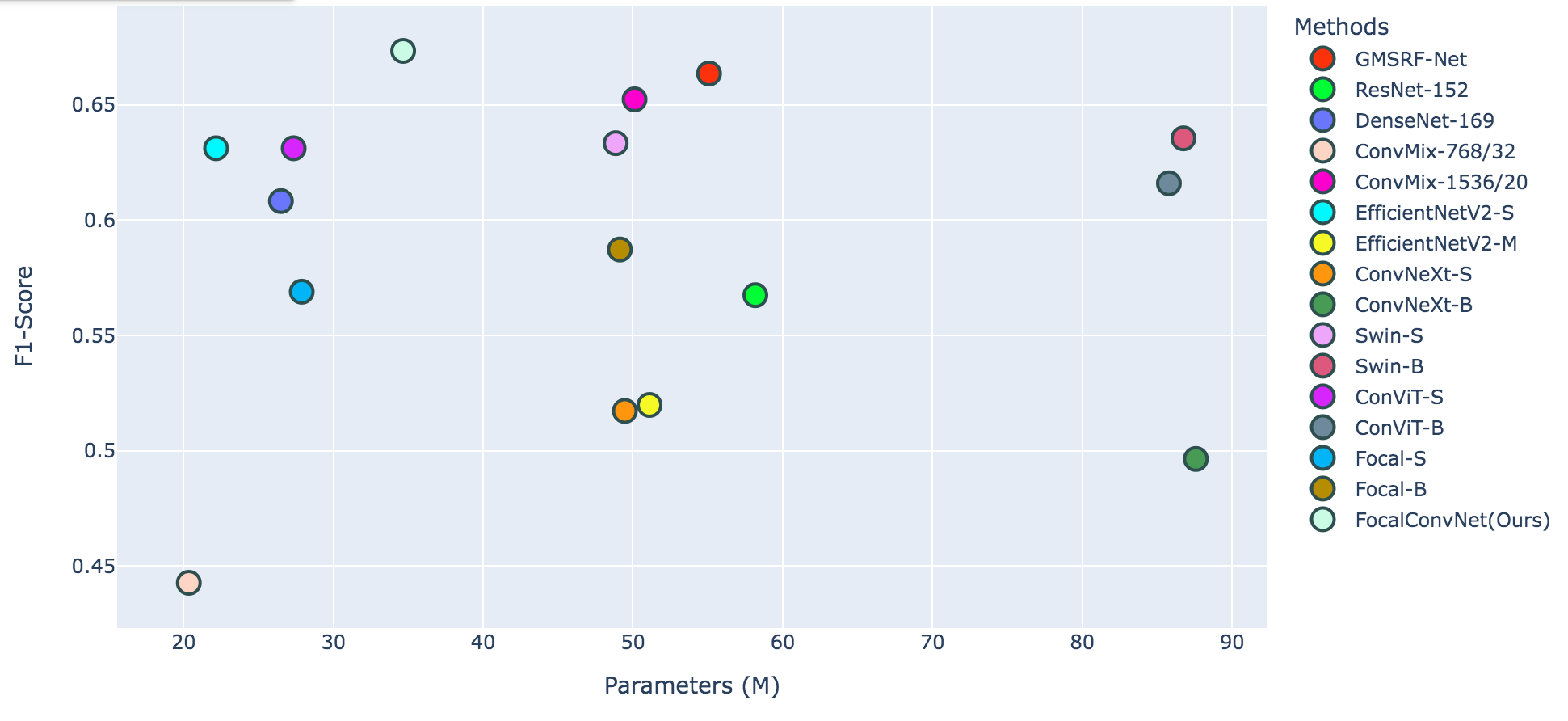}
    \caption{\textbf{Plot of F1-score vs parameters for each method}. Here, we can observe the trade-off between number of parameters and performance for all methods.}
    \label{fig:f1vsparams}
    \vspace{-5mm}
\end{figure*}

\begin{table}[!t]
\centering
\footnotesize
\caption{Comparison of FocalConvNet with baselines in terms of computational requirement (M denotes million) and throughput(number of images processed per second).}
\begin{tabular}{@{}c|c|c|c|c@{}}
\toprule
\textbf{Method} & {\bf Paramaters} & {\bf Year} & {\bf GFLOPs} & {\bf Throughput}\\ \hline
GMSRF-Net~\cite{srivastava2021gmsrf} & 55.08 M & 2021 & 160.88 & 7.60 \\  \hline
ResNet-152~\cite{he2016deep} & 58.17 M & 2016 & 11.58 & 53.16 \\ \hline
DenseNet-169~\cite{huang2017densely} & 26.5 M & 2017 & 7.82 & 130.79 \\ \hline
ConvMix-768/32~\cite{trockman2022patches} & 20.35 M & 2022 & 20.88 & 49.34 \\ \hline 
ConvMix-1536/20~\cite{trockman2022patches} & 50.11 M & 2022 & 51.36 & 19.20 \\ \hline 
EfficientNetV2-S~\cite{tan2021efficientnetv2} & 22.17 M & 2021 & 2.97 & 65.23 \\ \hline
EfficientNetV2-M~\cite{tan2021efficientnetv2} & 51.11 M & 2021 & 6.24 & 59.17 \\ \hline
ConvNeXt-S~\cite{liu2022convnet} & 49.46 M & 2022 & 8.70 & 69.53 \\ \hline
ConvNeXt-BS~\cite{liu2022convnet} & 87.58 M & 2022 & 15.38 & 50.36 \\ \hline 
Swin-S~\cite{dong2021cswin} & 48.85 M & 2021 & 8.52 & 102.07 \\ \hline
Swin-B~\cite{dong2021cswin} & 86.75 M & 2021 & 15.14 & 83.67 \\ \hline
ConViT-S~\cite{d2021convit} & 27.35 M & 2021 & 5.35 & 136.66 \\ \hline
ConViT-B~\cite{d2021convit} & 85.78 M & 2021 & 16.8 & 45.56 \\ \hline
Focal-S~\cite{yang2022focal} & 27.89 M & 2022 & 8.70 &  134.98 \\ \hline
Focal-B~\cite{yang2022focal} & 49.13 M & 2022 & 15.40 & 123.27 \\ \hline
FocalConvNet(Ours) & 34.66 M & 2022 & 5.23 & \textbf{148.02} \\ 
\bottomrule
\end{tabular}
\label{tab:result2}
\end{table}

\section{Result and Discussion}
In this section, we compare our proposed FocalConvNet with other \ac{SOTA} baselines on the Kvasir-Capsule dataset. Table~\ref{tab:result1} shows the quantitative comparison where both the \enquote{weighted} and \enquote{macro} averaging strategies are used while calculating F1-score, precision and recall. Additionally, we calculate the \ac{MCC} achieved by each method. From Table~\ref{tab:result1} we can observe that FocalConvNet attains a 0.98\% and 2.78\% improvement in F1-score and accuracy, respectively, over second-best performer GMSRF-Net. The \ac{MCC} reported by FocalConvNet is a 1.95\% increment over best performing ConViT-S. Additionally, we report the highest precision, recall, and F1-score of 0.2438, 0.2745, and 0.2178, respectively, when macro averaging is used. Consequently, it can be inferred that the multi-scale context guided visual modelling coupled with the inherent inductive bias of convolutions within our FocalConvNet is advantageous and powerful while extracting discriminative features.   

Table~\ref{tab:result2} compares the parameters, one billion floating-point operations per second (GFLOPs), and Throughput (images/second) of the proposed method with other baselines. It can be noted that FocalConvNet along with obtaining the highest weighted average F1-score, \ac{MCC} and accuracy, achieves the highest throughput of 148.02 despite having greater computational complexity than Focal-S, ConViT-S, EfficientNetV2-S, ConvMix-768/32 and DenseNet-169. Even though Focal-S and Focal-B use focal modulation as the key element in their architecture, it remains devoid of aggregation of local features which can otherwise be achieved by convolutional layers. Consequently, it gets outperformed by several other \ac{CNN} and transformer-based methods. Both the variants of Swin and ConViT report significant performance underscoring the power of vision transformers to effectively learn global and local interactions.
Figure~\ref{fig:f1vsthrough} illustrates the F1-score vs throughput observed by each method. Here, it can be ascertained that our proposed method is superior in performance when both F1-score and inference time is considered. An identical trend is followed when recall and \ac{MCC} are plotted against throughput for all methods. GMSRF-Net utilizing its global multi-scale fusion mechanism learns strong representations boosted by the global context and reports comparable results. Nevertheless, its high computation requirements (160.88 GFLOPs) and lowest throughput (7.60 images/second) are not favourable in a real-time setting.

This property of the FocalConvNet can be attributed to focal modulation and the lightweight convolutional block. Focal modulation is proficient in learning visual token interactions while serving as a light-weight replacement for self-attention. The convolutional block leverages $1\times1$ convolutions and depth-wise convolutions to serve as a proxy for computationally heavy $3\times3$ convolutions. Thus, in medical image classification where data tends to be scarce or imbalanced in nature, our FocalConvNet can serve as an effective and efficient baseline in future. Moreover, in Figure~\ref{fig:f1vsparams} it can be seen that in the performance vs complexity graph, FocalConvNet shows a decent trade-off.

\section{Conclusion}
In this paper, we propose a novel lightweight and swift medical classification architecture for real-time anatomical and luminal findings (pathological and mucosal view) classification in video capsule endoscopy. The proposed FocalConvNet leverages the learning bias in convolutions and mixes it with the global and local representation learning power of focal modulation to give favourable performance on Kvasir-Capsule. Combination of focal modulation and lightweight convolutions enable the FocalConvNet to not only outperform other \ac{SOTA} baselines, but also report the highest throughput. Consequently, our proposed method can be used in real-time analysis of video capsule endoscopy. Thus, reducing the effort required for manual inspection. We also use several \ac{SOTA} baselines and benchmark them on Kvasir-Capsule to streamline further research in this area. In the future, we plan to leverage generative models to abate the class imbalance and further increase the performance. 



\bibliographystyle{IEEEtran}
\bibliography{references} 

\begin{thebibliography}{10}
\providecommand{\url}[1]{#1}
\csname url@samestyle\endcsname
\providecommand{\newblock}{\relax}
\providecommand{\bibinfo}[2]{#2}
\providecommand{\BIBentrySTDinterwordspacing}{\spaceskip=0pt\relax}
\providecommand{\BIBentryALTinterwordstretchfactor}{4}
\providecommand{\BIBentryALTinterwordspacing}{\spaceskip=\fontdimen2\font plus
\BIBentryALTinterwordstretchfactor\fontdimen3\font minus
  \fontdimen4\font\relax}
\providecommand{\BIBforeignlanguage}[2]{{%
\expandafter\ifx\csname l@#1\endcsname\relax
\typeout{** WARNING: IEEEtran.bst: No hyphenation pattern has been}%
\typeout{** loaded for the language `#1'. Using the pattern for}%
\typeout{** the default language instead.}%
\else
\language=\csname l@#1\endcsname
\fi
#2}}
\providecommand{\BIBdecl}{\relax}
\BIBdecl

\bibitem{iddan2000wireless}
G.~Iddan, G.~Meron, A.~Glukhovsky, and P.~Swain, ``Wireless capsule
  endoscopy,'' \emph{Nature}, vol. 405, no. 6785, p. 417, 2000.

\bibitem{lee2019natural}
S.~Y. Lee, J.~Y. Lee, Y.~J. Lee, and K.~S. Park, ``Natural elimination of a
  video capsule after retention for 1 year in a patient with small bowel crohn
  disease: A case report,'' \emph{Medicine}, vol.~98, no.~43, 2019.

\bibitem{suman2017detection}
S.~Suman, A.~S. Malik, K.~Pogorelov, M.~Riegler, S.~H. Ho, I.~Hilmi, K.~L. Goh
  \emph{et~al.}, ``Detection and classification of bleeding region in wce
  images using color feature,'' in \emph{Proceedings of the 15th International
  Workshop on Content-Based Multimedia Indexing}, 2017, p.~17.

\bibitem{smedsrud2021kvasir}
P.~H. Smedsrud, V.~Thambawita, S.~A. Hicks, H.~Gjestang, O.~O. Nedrejord,
  E.~N{\ae}ss, H.~Borgli, D.~Jha, T.~J.~D. Berstad, S.~L. Eskeland
  \emph{et~al.}, ``Kvasir-capsule, a video capsule endoscopy dataset,''
  \emph{Scientific Data}, vol.~8, no.~1, pp. 1--10, 2021.

\bibitem{meerveld2017gastrointestinal}
G.-V. Meerveld, A.~C. Johnson, D.~Grundy \emph{et~al.}, ``Gastrointestinal
  physiology and function,'' \emph{Gastrointestinal pharmacology}, pp. 1--16,
  2017.

\bibitem{ENDOCAPSULE-10}
Olympus, ``The endocapsule 10 system,'' \textit{Olympus homepage},
  \url{https://www.olympus-europa.com/medical/en/Products-and-Solutions/Products/Product/ENDOCAPSULE-10-System.html},
  2013.

\bibitem{kaminski2010quality}
M.~F. Kaminski, J.~Regula, E.~Kraszewska, M.~Polkowski, U.~Wojciechowska,
  J.~Didkowska, M.~Zwierko, M.~Rupinski, M.~P. Nowacki, and E.~Butruk,
  ``Quality indicators for colonoscopy and the risk of interval cancer,''
  \emph{New England Journal of Medicine}, vol. 362, no.~19, pp. 1795--1803,
  2010.

\bibitem{yang2022focal}
J.~Yang, C.~Li, and J.~Gao, ``Focal modulation networks,'' \emph{arXiv preprint
  arXiv:2203.11926}, 2022.

\bibitem{krizhevsky2012imagenet}
A.~Krizhevsky, I.~Sutskever, and G.~E. Hinton, ``Imagenet classification with
  deep convolutional neural networks,'' \emph{Advances in neural information
  processing systems}, vol.~25, 2012.

\bibitem{simonyan2014very}
K.~Simonyan and A.~Zisserman, ``Very deep convolutional networks for
  large-scale image recognition,'' \emph{arXiv preprint arXiv:1409.1556}, 2014.

\bibitem{he2016deep}
K.~He, X.~Zhang, S.~Ren, and J.~Sun, ``Deep residual learning for image
  recognition,'' in \emph{Proceedings of the IEEE conference on computer vision
  and pattern recognition (CVPR)}, 2016, pp. 770--778.

\bibitem{huang2017densely}
G.~Huang, Z.~Liu, L.~Van Der~Maaten, and K.~Q. Weinberger, ``Densely connected
  convolutional networks,'' in \emph{Proceedings of the IEEE conference on
  computer vision and pattern recognition}, 2017, pp. 4700--4708.

\bibitem{wang2020deep}
J.~Wang, K.~Sun, T.~Cheng, B.~Jiang, C.~Deng, Y.~Zhao, D.~Liu, Y.~Mu, M.~Tan,
  X.~Wang \emph{et~al.}, ``Deep high-resolution representation learning for
  visual recognition,'' \emph{IEEE transactions on pattern analysis and machine
  intelligence}, vol.~43, no.~10, pp. 3349--3364, 2020.

\bibitem{9662196}
A.~Srivastava, D.~Jha, S.~Chanda, U.~Pal, H.~D. Johansen, D.~Johansen, M.~A.
  Riegler, S.~Ali, and P.~Halvorsen, ``{MSRF-Net}: {A Multi-Scale Residual
  Fusion Network for Biomedical Image Segmentation},'' \emph{IEEE Journal of
  Biomedical and Health Informatics}, vol.~26, no.~5, pp. 2252--2263, 2022.

\bibitem{srivastava2021gmsrf}
A.~Srivastava, S.~Chanda, D.~Jha, U.~Pal, and S.~Ali, ``{GMSRF-Net}: {An
  improved generalizability with global multi-scale residual fusion network for
  polyp segmentation},'' in \emph{Proceedings of the International conference
  on pattern recognition}, 2022.

\bibitem{hu2018squeeze}
J.~Hu, L.~Shen, and G.~Sun, ``Squeeze-and-excitation networks,'' in
  \emph{Proceedings of the IEEE conference on computer vision and pattern
  recognition}, 2018, pp. 7132--7141.

\bibitem{woo2018cbam}
S.~Woo, J.~Park, J.-Y. Lee, and I.~S. Kweon, ``Cbam: Convolutional block
  attention module,'' in \emph{Proceedings of the European conference on
  computer vision (ECCV)}, 2018, pp. 3--19.

\bibitem{chollet2017xception}
F.~Chollet, ``Xception: Deep learning with depthwise separable convolutions,''
  in \emph{Proceedings of the IEEE conference on computer vision and pattern
  recognition}, 2017, pp. 1251--1258.

\bibitem{hua2018pointwise}
B.-S. Hua, M.-K. Tran, and S.-K. Yeung, ``Pointwise convolutional neural
  networks,'' in \emph{Proceedings of the IEEE Conference on Computer Vision
  and Pattern Recognition}, 2018, pp. 984--993.

\bibitem{liu2022convnet}
Z.~Liu, H.~Mao, C.-Y. Wu, C.~Feichtenhofer, T.~Darrell, and S.~Xie, ``A convnet
  for the 2020s,'' \emph{arXiv preprint arXiv:2201.03545}, 2022.

\bibitem{trockman2022patches}
A.~Trockman and J.~Z. Kolter, ``Patches are all you need?'' \emph{arXiv
  preprint arXiv:2201.09792}, 2022.

\bibitem{dosovitskiy2020image}
A.~Dosovitskiy, L.~Beyer, A.~Kolesnikov, D.~Weissenborn, X.~Zhai,
  T.~Unterthiner, M.~Dehghani, M.~Minderer, G.~Heigold, S.~Gelly \emph{et~al.},
  ``An image is worth 16x16 words: Transformers for image recognition at
  scale,'' \emph{arXiv preprint arXiv:2010.11929}, 2020.

\bibitem{touvron2021training}
H.~Touvron, M.~Cord, M.~Douze, F.~Massa, A.~Sablayrolles, and H.~J{\'e}gou,
  ``Training data-efficient image transformers \& distillation through
  attention,'' in \emph{Proceedings of the International Conference on Machine
  Learning}.\hskip 1em plus 0.5em minus 0.4em\relax PMLR, 2021, pp.
  10\,347--10\,357.

\bibitem{yu2021metaformer}
W.~Yu, M.~Luo, P.~Zhou, C.~Si, Y.~Zhou, X.~Wang, J.~Feng, and S.~Yan,
  ``Metaformer is actually what you need for vision,'' \emph{arXiv preprint
  arXiv:2111.11418}, 2021.

\bibitem{d2021convit}
S.~d’Ascoli, H.~Touvron, M.~L. Leavitt, A.~S. Morcos, G.~Biroli, and
  L.~Sagun, ``Convit: Improving vision transformers with soft convolutional
  inductive biases,'' in \emph{Proceedings of the International Conference on
  Machine Learning}, 2021, pp. 2286--2296.

\bibitem{arnab2021vivit}
A.~Arnab, M.~Dehghani, G.~Heigold, C.~Sun, M.~Lu{\v{c}}i{\'c}, and C.~Schmid,
  ``Vivit: A video vision transformer,'' in \emph{Proceedings of the IEEE/CVF
  International Conference on Computer Vision}, 2021, pp. 6836--6846.

\bibitem{dong2021cswin}
X.~Dong, J.~Bao, D.~Chen, W.~Zhang, N.~Yu, L.~Yuan, D.~Chen, and B.~Guo,
  ``Cswin transformer: A general vision transformer backbone with cross-shaped
  windows,'' \emph{arXiv preprint arXiv:2107.00652}, 2021.

\bibitem{rao2021dynamicvit}
Y.~Rao, W.~Zhao, B.~Liu, J.~Lu, J.~Zhou, and C.-J. Hsieh, ``Dynamicvit:
  Efficient vision transformers with dynamic token sparsification,''
  \emph{Advances in neural information processing systems}, vol.~34, 2021.

\bibitem{wang2021pyramid}
W.~Wang, E.~Xie, X.~Li, D.-P. Fan, K.~Song, D.~Liang, T.~Lu, P.~Luo, and
  L.~Shao, ``Pyramid vision transformer: A versatile backbone for dense
  prediction without convolutions,'' in \emph{Proceedings of the IEEE/CVF
  International Conference on Computer Vision}, 2021, pp. 568--578.

\bibitem{fan2021multiscale}
H.~Fan, B.~Xiong, K.~Mangalam, Y.~Li, Z.~Yan, J.~Malik, and C.~Feichtenhofer,
  ``Multiscale vision transformers,'' in \emph{Proceedings of the IEEE/CVF
  International Conference on Computer Vision}, 2021, pp. 6824--6835.

\bibitem{dai2021coatnet}
Z.~Dai, H.~Liu, Q.~V. Le, and M.~Tan, ``Coatnet: Marrying convolution and
  attention for all data sizes,'' \emph{Advances in Neural Information
  Processing Systems}, vol.~34, pp. 3965--3977, 2021.

\bibitem{khan2021transformers}
S.~Khan, M.~Naseer, M.~Hayat, S.~W. Zamir, F.~S. Khan, and M.~Shah,
  ``Transformers in vision: A survey,'' \emph{ACM Computing Surveys (CSUR)},
  2021.

\bibitem{jha2022machine}
D.~Jha, ``Machine learning-based classification, detection, and segmentation of
  medical images,'' \emph{PhD thesis}, 2022.

\bibitem{tan2021efficientnetv2}
M.~Tan and Q.~Le, ``Efficientnetv2: Smaller models and faster training,'' in
  \emph{Proceedings of the International Conference on Machine Learning}, 2021,
  pp. 10\,096--10\,106.

\bibitem{jha2021comprehensive}
D.~Jha, S.~Ali, S.~Hicks, V.~Thambawita, H.~Borgli, P.~H. Smedsrud,
  T.~de~Lange, K.~Pogorelov, X.~Wang, P.~Harzig \emph{et~al.}, ``A
  comprehensive analysis of classification methods in gastrointestinal
  endoscopy imaging,'' \emph{Medical image analysis}, vol.~70, p. 102007, 2021.

\end{thebibliography}

\ifCLASSOPTIONcaptionsoff
  \newpage
\fi
\vfill
\end{document}